\documentclass[twocolumn]{aastex62}

\newcommand{\ca}{\ensuremath{^{48}\mathrm{Ca}}}
\newcommand{\ti}{\ensuremath{^{50}\mathrm{Ti}}}
\newcommand{\msun}{\ensuremath{M_\sun}}

\graphicspath{{./}{figures/}}

\received{June 7, 2019}
\revised{July 31, 2019}
\accepted{August 1, 2019}

\shorttitle{ECSNe in GCE}
\shortauthors{Jones, C\^ot\'e, R\"opke \& Wanajo}

\begin{document}

\title{A New Model for Electron-Capture Supernovae in Galactic Chemical
Evolution}

\correspondingauthor{Samuel Jones}
\email{swjones@lanl.gov}

\author[0000-0003-3970-1843]{Samuel Jones}
\affil{X Computational Physics (XCP) Division and Center for Theoretical
	Astrophysics (CTA), Los Alamos National Laboratory, \\
Los Alamos, New Mexico 87545, USA}

\author[0000-0002-9986-8816]{Benoit C\^ot\'e}
\affiliation{Konkoly Observatory, Research Centre for Astronomy and Earth Sciences, Hungarian
Academy of Sciences, Konkoly Thege Miklos ut 15-17, H-1121 Budapest, Hungary}
\affiliation{National Superconducting Cyclotron Laboratory, Michigan State University, East
Lansing, MI 48824, USA}
\affiliation{Joint Institute for Nuclear Astrophysics - Center for the Evolution of the
Elements, USA}

\author[0000-0002-4460-0097]{Friedrich~K.~R\"opke}
\affiliation{Institut f{\"u}r Theoretische Astrophysik, Zentrum f\"ur Astronomie der Universit\"at Heidelberg, Philosophenweg 12, 69120 Heidelberg, Germany}
\affiliation{Heidelberg Institute for Theoretical Studies, Schloss-Wolfsbrunnenweg 35, 69118 Heidelberg, Germany}

\author[0000-0002-4759-7794]{Shinya Wanajo}
\affiliation{Max Planck Institute for Gravitational Physics (Albert Einstein
Institute), Am M\"uhlenberg 1, Potsdam-Golm, D-14476, Germany}
\affil{Department of Engineering and Applied Sciences, Sophia University, Chiyoda-ku, Tokyo 102-8554, Japan}
\affil{iTHEMS Research Group, RIKEN, Wako, Saitama 351-0198, Japan}



\begin{abstract}

	We examine the contribution of electron-capture supernovae (ECSNe),
	low-mass SNe from collapsing Fe cores (FeCCSNe), and
	rotating massive stars to the chemical composition of the Galaxy.
	Our model includes contributions to chemical evolution from both
	thermonuclear ECSNe (tECSNe) and gravitational collapse ECSNe (cECSNe).
	We show that if ECSNe are predominantly gravitational collapse SNe but
	about 15\% are partial thermonuclear explosions, the model is able to
	reproduce the solar abundances of several important and problematic
	isotopes including \ca, \ti~and $^{54}$Cr together with $^{58}$Fe,
	$^{64}$Ni, $^{82}$Se and $^{86}$Kr and several of the Zn--Zr isotopes.
	A model in which no cECSNe occur, only tECSNe with low-mass
	FeCCSNe or rotating massive stars, proves also very successful
	at reproducing the solar abundances for these isotopes.
	Despite the small mass range for the progenitors of ECSNe and
	low-mass FeCCSNe, the large production factors suffice for the solar
	inventory of the above isotopes.
	Our model is compelling because it introduces no new tensions with the
	solar abundance distribution for a Milky Way model -- only tending to
	improve the model predictions for several isotopes.
	The proposed astrophysical production model thus provides a natural and
	elegant way to explain one of the last uncharted territories on the
	periodic table of astrophysical element production.

\end{abstract}


\section{Introduction} \label{sec:intro}

Recent progress in observations \citep{Abbott2017a} and modeling
\citep{Cowan2019a} lend strong support to neutron star mergers as the main
astrophysical site of the $r$~process. Its counterpart neutron-capture processes
-- the weak and main $s$~processes, operating in massive and AGB stars
\citep{Busso1999a,Kaeppeler2011a}, respectively -- produce familiar abundance
patterns that have been studied for several decades now with quite some rigour.
The production of the lighter elements (He--Fe) in AGB and massive stars
\citep{HerwigAGB2005,Nomoto2013a}, and in thermonuclear and core-collapse
supernovae \citep{Seitenzahl2013a,Woosley2002} have also been extensively
studied.

\begin{figure*}
	\includegraphics[width=0.98\textwidth]{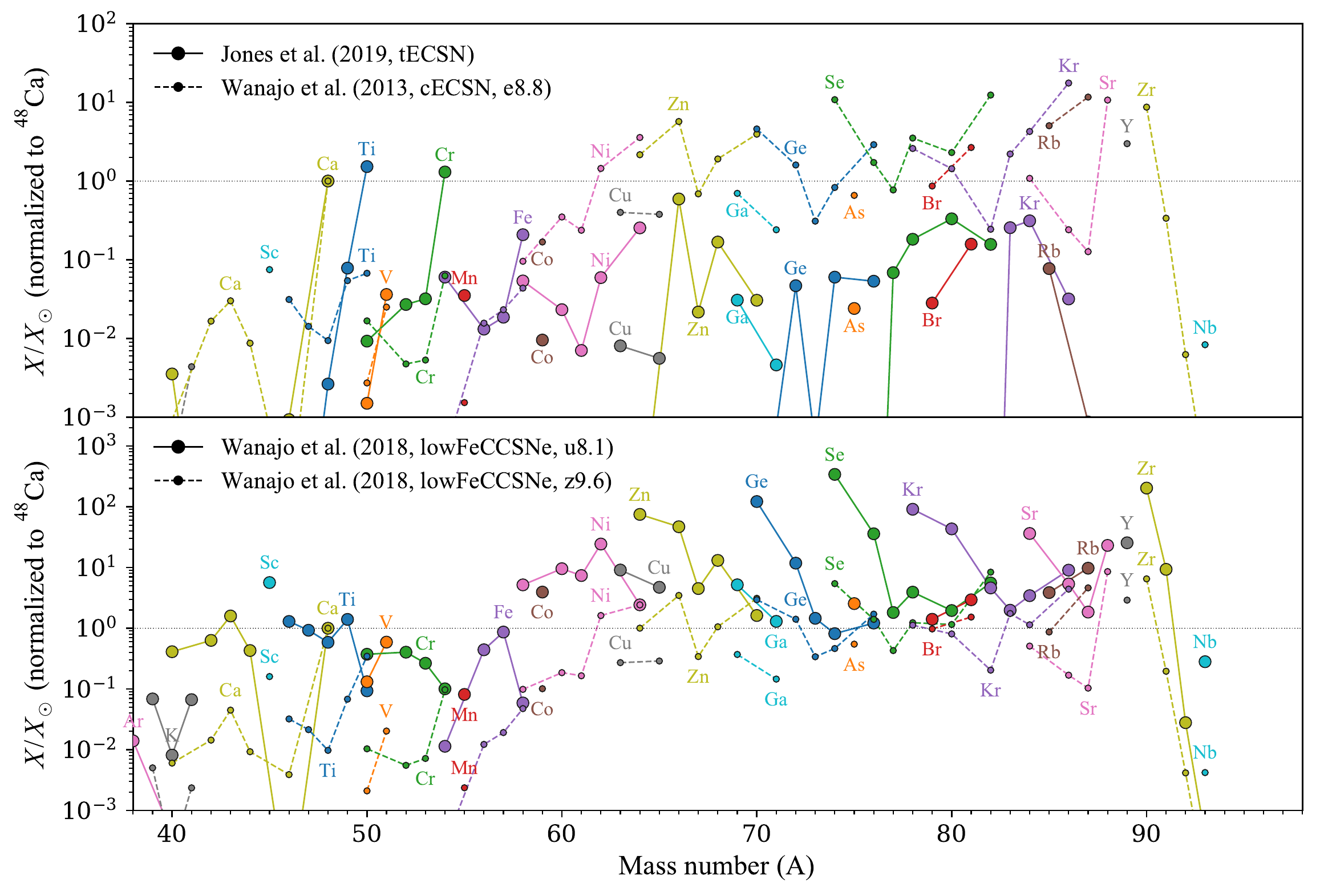}
	\caption{
		\textit{Top Panel:} Comparison of ejected compositions from the
		thermonuclear ECSN model of \citet[solid lines]{Jones2019a}
		and the gravitational collapse e8.8 ECSN model of \citet[dashed
		lines]{Wanajo2013b}. \textit{Bottom Panel:} Comparison of
		ejected compositions from the u8.1 (solid lines) and z9.6
		(dashed lines) low-mass FeCCSN models of \cite{Wanajo2018a}. All
		compositions have first been normalized to the solar isotopic
		composition, and then re-normalized to shift $^{48}$Ca to 1.0 by
		dividing by the overproduction factor of $^{48}$Ca,
		$X_\mathrm{ejected}(^{48}\mathrm{Ca})/X_\sun(^{48}\mathrm{Ca})$.
		The overproduction factors for $^{48}$Ca were 19, 0.40 and 19
		for e8.8, u8.1 and z9.6, respectively. The equivalent value for
		the G14a tECSN model from \citet{Jones2019a} is
		$1.40\times10^{4}$ without an envelope, or $1.58\times10^3$ if
		one were to assume that the progenitor was an 8.8~\msun~star.
	}
	\label{fig:both_yields}
\end{figure*}

Although our picture of the astrophysical production of nuclear species
improves, the origins of a number of isotopes still remain a mystery. One of
the remaining blemishes on this nuclear charts are the neutron-rich
isotopes \ca, \ti, $^{54}$Cr and isotopes of the elements in the Zn--Zr region.
Their production requires special conditions that are reached in explosive
thermonuclear burning in high-density material in which electron-captures
produce a low $Y_\mathrm{e}$ \citep{Meyer1996a}. This has been discussed by
\citet{Woosley1997a} in the context of thermonuclear explosion of very high
density carbon/oxygen white dwarfs (CO WDs); however, stellar evolution theory
does not support the existence of such objects.

The required nucleosynthetic conditions can, however, be reached in so-called
electron-capture supernovae (ECSNe) in massive degenerate oxygen neon (ONe)
cores that form in the final evolutionary stages of stars in the mass range of
about 8--10~\msun, in-between AGB stars and massive stars \citep{Nomoto1984a,
Nomoto1987}. For these, two explosion scenarios are discussed: a collapse into a
neutron star \citep[NS; cECSN;][]{Wanajo2011} and a (partial) thermonuclear
disruption \citep[tECSN;][]{Jones2016a} leaving behind an ONeFe white dwarf (WD)
remnant.  Though recent attempts have been made to predict which explosion
occurs in nature \citep{Jones2019a,Leung2019a}, there remain outstanding
uncertainties that make it extremely challenging.
It should also be noted that a recent chemical evolution study
\citep{Prantzos2018a} demonstrates that the weak $s$ process in rotating massive
stars \citep{limongichieffi18} can be the predominant source of the elements in
the Zn-Zr region.

Here, we present an argument that the thermonuclear explosion channel indeed
occurs in nature, probably at a low rate, and is primarily responsible for the
solar inventory of \ca, \ti, $^{54}$Cr and provides a substantial contribution
to $^{58}$Fe, $^{64}$Ni and $^{66,68}$Zn.  We show that this model avoids
inconsistencies in the production factors of other isotopes. We further develop
our model to include some fraction of ECSNe that collapse into NSs and combine
the ECSN yields with those from low-mass Fe-core explosions
\citep[FeCCSNe,][]{Wanajo2018a} or rotating massive stars
\citep{limongichieffi18}.

\begin{figure}
	\includegraphics[width=\linewidth]{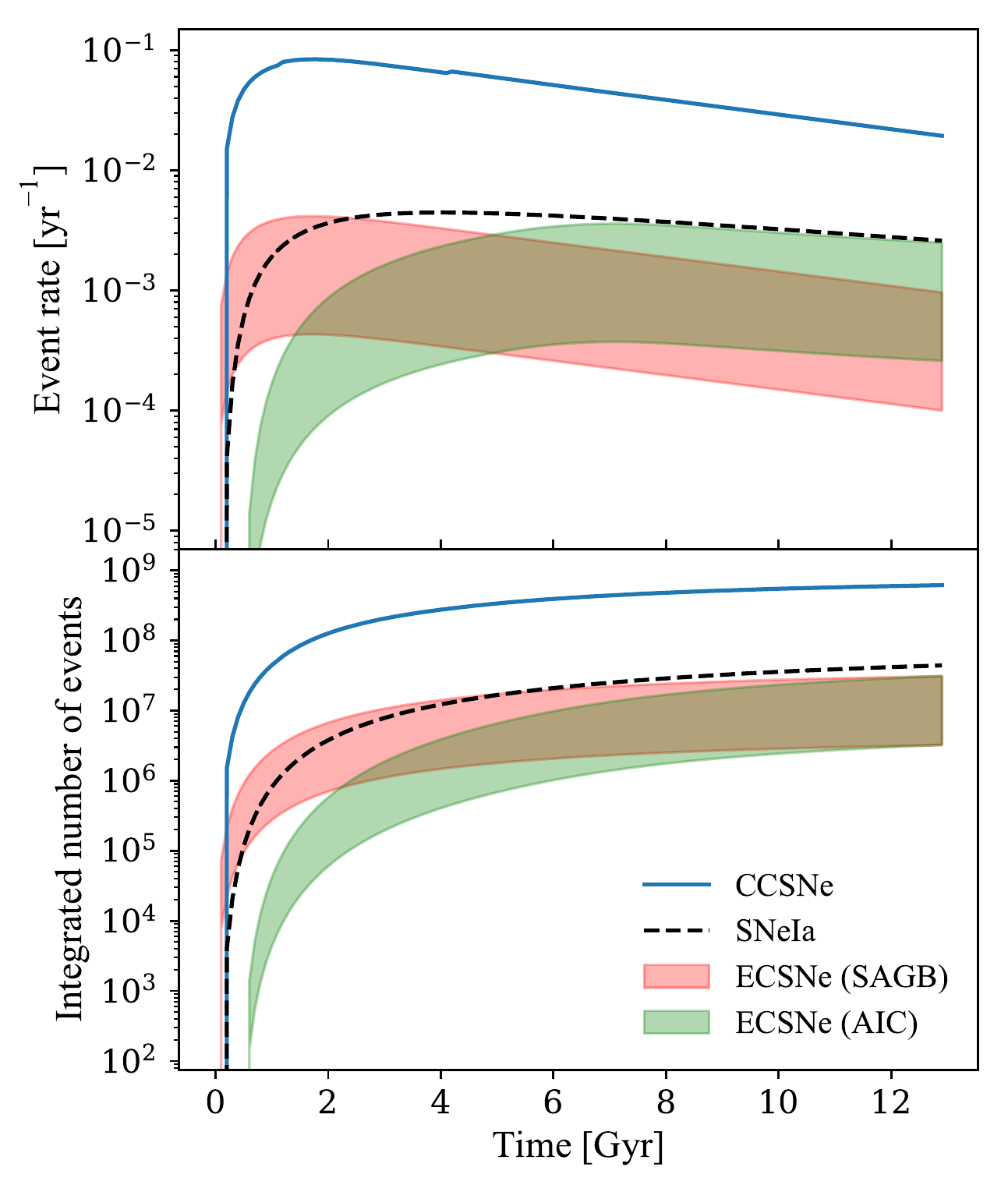}
	\caption{
		\emph{Top Panel:} Rate of SN events over the course of the
	simulation. We assume the Sun forms after 8.5\,Gyr of evolution.
		\emph{Bottom Panel:} Integrated number of SN events. The broad
		bands show the predicted contribution of ECSNe, assuming the
		progenitors are single stars (SAGB, red bands) or accreting ONe
		WDs (AIC, green bands). For the red bands, the lower and upper
		values correspond to 0.5\,\% and 5\,\% the rate of CCSNe. The
		former is the rate needed for thermonuclear ECSNe to reproduce
		the abundance of $^{48}$Ca, and the latter is the rate adopted
		for the contribution of gravitational collapse ECSNe shown in
		Fig.~\ref{fig:gce_both}b. Assuming accreting ONe WDs for the
		progenitors of ECSNe, those rates needed to be increased by
		40\,\% in order to recover the same number of ECSNe by the end
		of the simulations (see green band in Bottom Panel).
	}
	\label{fig:gce_events}
\end{figure}

\section{Galactic chemical evolution models}

\subsection{Code description and ingredients}

To bring the nucleosynthesis yields into a galactic context, we use the
open-source two-zone chemical evolution model \texttt{OMEGA+}
(\citealt{cote18}). The adopted Milky Way setup is described in
\cite{cote19}\footnote{
	\url{https://github.com/becot85/JINAPyCEE/blob/master/DOC/OMEGA\%2B_Milky_Way_model.ipynb}
	}
	and allows to reproduce a variety of observational constraints including
	the current star formation rate, gas inflow rate, star-to-gas mass
	ratio, and core-collapse and Type~Ia supernova rates (CCSN, SNIa).  We
	use mass- and metallicity-dependent yields for both low- and
	intermediate-mass stars (\citealt{cristallo15}) and for massive stars
	(\citealt{limongichieffi18}, either rotating or non-rotating models).
	For each stellar population formed throughout the galactic chemical
	evolution (GCE) calculations, we fold the yields with the initial mass
	function of \cite{kroupa01}. We use the delayed-detonation N100 model of
	\cite{Seitenzahl2013a} for the yields of SNe~Ia, and distribute them
	within each stellar population following a delay-time distribution (DTD)
	function in the form of $t^{-1}$ (\citealt{maoz14}, see
	\citealt{ritter_sygma} for implementation details). In total, we
	generate $\sim$\,10$^{-2}$ and $\sim$\,10$^{-3}$ CCSN and SNIa per unit
	of stellar mass ($M_\odot$) formed, respectively.

\subsection{Inclusion of electron-capture supernovae in GCE model}

We have considered two sources of yields for ECSNe: those by \citet{Jones2019a}
for tECSNe and those by \citet{Wanajo2013b} for cECSNe, with ejecta masses of
0.95~\msun~and 0.011~\msun~respectively. The abundance distributions for the two
yield sets are compared in Fig.~\ref{fig:both_yields}.

The population synthesis calculations in \citet{Jones2019a} based on models from
\citet{Ruiter2019a} suggest that the most common evolutionary channel producing
ECSNe is from stars evolving directly towards explosion in binary systems (2.8\%
of CCSN rate), as opposed to accreting ONe WDs (so-called accretion-induced
collapse, AIC, 0.36\% of CCSN rate) or isolated single SAGB stars (0.15\%).
However, it is not clear whether this is the case at all metallicities
\citep[][and references therein]{Doherty2015}. Further population synthesis
studies would be highly desirable, in which the underlying assumptions and
models are updated and based on the most recent single and binary star models
\citep{Tauris2015a,Poelarends2017a,Siess2018a}.

There are hence two assumptions one could make for the ECSN DTD that bracket the
range of possible DTDs: the SAGB channel DTD would be CCSN-like (i.e., with the
stellar lifetimes and no further time delay), and the AIC channel would follow
more of a single-degenerate SN~Ia DTD.  We assume that the DTD for stars
evolving directly to an ECSN in a binary will be similar to the CCSN-like DTD
because no accretion phenomena are involved.

The event rate and integrated number of events as a function of time in the GCE
simulations is shown in Fig.~\ref{fig:gce_events}. The red (green) bands show
these quantities for ECSNe assuming a CCSN-like (AIC-like) DTD.
The AIC DTD was constructed based on the results of \citet{ruiter2009a}.
In this study
the red (CCSN-like) band was used. The lower limit is at 0.5\% of the CCSN rate,
and the upper limit is at 5\% of the CCSN rate. These limits represent the
range of ECSN rates that we use for this work.
A tECSN rate that is no more than 0.5\% of the CCSN rate is needed to reproduce
all of the solar \ca~for a CCSN-like DTD, which increases to 0.7\% for and
AIC-like DTD.

\section{Results}
\subsection{Thermonuclear ECSN\lowercase{e} as the origin of \ca, \ti,
$^{54}$C\lowercase{r}, $^{58}$F\lowercase{e}, $^{64}$N\lowercase{i} and
$^{66,68}$Z\lowercase{n}}
\label{sec:ca_results}
In this section we demonstrate that tECSNe are able to account for the solar
inventory of \ca~and several other neutron-rich isotopes without introducing new
tensions.

\begin{figure*}
	\includegraphics[width=0.98\textwidth]{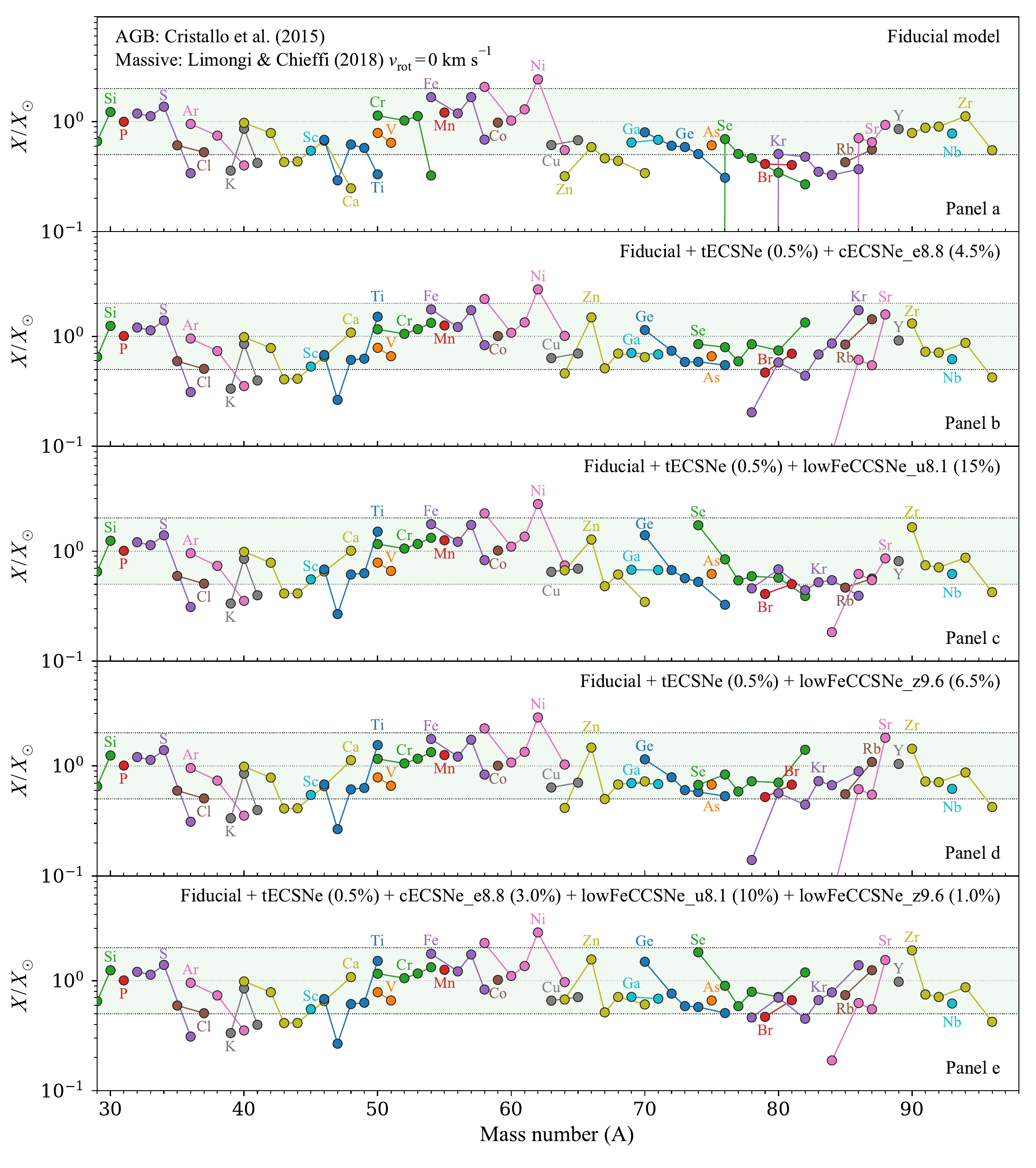}
	\caption{
		Galaxy model composition relative to solar when the Sun forms.
		\textit{Panel a:} Fiducial model without ECSN and low-mass Fe
		CCSN (low-FeCCSNe). \textit{Panel b:} Predictions assuming a
		combination of thermonuclear SNe (tECSNe) and gravitational
		collapse ECSNe (cECSNe). \textit{Panel c:} Combination of tECSNe
		and lowFeCCSNe (u8.1). \textit{Panel d:} Combination of tECSNe
		and lowFeCCSNe (z9.6). \textit{Panel e:} Combination of tECSNe,
		cECSNe, and lowFeCCSNe (u8.1 and z9.6). The percentages in
		parenthesis in the panel legend represent the rate of the
		considered site in percentage of the CCSN rate. Panels B through
		E show three additional isotopes ($^{74}$Se, $^{78}$Kr, and
		$^{84}$Sr, which are $p$-isotopes) compared to the fiducial
		prediction shown in Panel A. Those isotopes are not present in
		the yields of \cite{cristallo15} and \cite{limongichieffi18}.
	}
	\label{fig:gce_both}
\end{figure*}

In Fig.~\ref{fig:gce_both} we plot the composition of our Milky Way models
relative to the solar composition at the time when the Sun forms.
Fig.~\ref{fig:gce_both}a shows our fiducial model in which no ECSNe or rotating
massive stars were included at all. We note the underproduction of \ca, \ti,
$^{54}$Cr and several isotopes in the Zn--Zr region.  We also note at this time
that $^{62}$Ni is already overproduced in our fiducial model by more than a
factor of two. This comes from the s-process yields we are using
\citep{limongichieffi18}.

If one attempts to explain the solar \ca~with cECSNe, one would not only need a
much higher ECSN rate than is expected ($\sim 65\%$), but at such a high rate
many of the light trans-Fe isotopes would be overproduced by up to an order of
magnitude (Fig.~\ref{fig:both_yields}). Furthermore, the large ratio of \ca~to
both \ti~and $^{54}$Cr in cECSNe is not compatible with the solar abundances.
Therefore if all ECSNe collapse into NSs, an additional source of these isotopes
would be required.

A model where tECSNe (cECSNe) have been assumed to occur at 0.5\% (4.5\%) of the
CCSN rate with a CCSN-like DTD on top of our fiducial model is shown in
Fig.~\ref{fig:gce_both}b. The addition of tECSNe affects only \ca, $^{50}$Ti,
$^{54}$Cr, $^{58}$Fe, $^{64}$Ni and $^{66,68}$Zn, and all of these isotopes
match the solar abundances better when tECSN yields are included, with the
former three being the most markedly improved. The rate estimate  of 0.5\%
(0.7\%) of the CCSN rate for the CCSN-like (AIC-like) DTD is similar to but
lower than the simpler estimate by \citet{Jones2019a}.

We note that while the model in Fig.~\ref{fig:gce_both}b reproduces well the
solar abundance of \ca, \ti~and $^{54}$Cr, \ti~and $^{54}$Cr are now slightly
overproduced in the best-fit model for \ca~(though within a factor of two).  The
$Y_\mathrm{e}$ of \ca~ is 0.417 and that of \ti~is 0.44, this hints at \ca~being
produced in slightly higher density conditions, or at least lower
$Y_\mathrm{e}$. There are still uncertainties in the hydrodynamic tECSN
simulations that could therefore affect the \ca/\ti~ratio in the ejecta.  For
example, the ratio is sensitive to the weak reaction rates used in the
nucleosynthesis calculations \citep{Jones2019a}. Moreover, the ratio is greater
than unity in the bound remnants, indicating that if slightly more of the lower
$Y_\mathrm{e}$ material were ejected the \ca/\ti~ratio might better match the
solar one. If the expansion time scale of the WD as the deflagration burns
through it were slightly longer, the simulation might also result in a more
favourable \ca/\ti~ratio. This uncertainty is related to not knowing the precise
ignition conditions of the ONe core (central density and initial ignition
geometry).

\subsection{A new model for GCE}

In this section we present models in which different combinations of tECSNe,
cECSNe and low-mass FeCCSNe occur. We demonstrate that such models -- in
particular one where all three types of SNe occur -- are quite successful at
reproducing the Solar composition when applied in a GCE code, especially for
several challenging isotopes.

\subsubsection{The role of tECSNe}

The models all assume a CCSN-like DTD for ECSNe with tECSNe occurring at 0.5\%
of the CCSN rate in order to match the solar abundance of \ca~by the time the
Sun forms. It is only by the inclusion of tECSNe that the \ca, \ti~and $^{54}$Cr
abundances are simultaneously and satisfactorily explained (as discussed in
Section~\ref{sec:ca_results}).

\subsubsection{cECSN compliment}

As \citet{Wanajo2011} have shown, cECSNe are a promising site for production of
the problematic region Zn--Zr (Fig.~\ref{fig:both_yields}), which is
underproduced when we include tECSNe at the necessary rate to match \ca.
Addionally, the diluted H/He envelopes of cECSNe do not contribute much
isotopes below $A = 48$, either. cECSNe therefore provide quite a good
compliment to tECSNe and together exhibit performance favourable to only
including one or the other. Indeed, when we include cECSNe at 4.5\% of the CCSN
rate in addition to tECSNe at 0.5\% of the CCSN rate (Fig.~\ref{fig:gce_both}b),
the whole picture looks much improved over the case where no ECSNe are included
(Fig.~\ref{fig:gce_both}a).

One might imagine that if the tECSN ejecta would reach only slightly lower
$Y_\mathrm{e}$ then more Zn--Zr would be produced (e.g. Fig.~1 of
\citealp{Jones2019a}), the \ca/\ti~ratio would fit even better and one may not
need cECSNe at all. However, the most neutron-rich isotopes such as $^{82}$Se
and $^{86}$Kr are more naturally fit with cECSNe, which exhibits higher
overproduction relative to the more proton-rich isotopes of Se and Kr, than
tECSNe.

\subsubsection{Low-mass FeCCSNe: is there a need for cECSNe at all?}

\citet{Wanajo2018a} have reported nucleosynthesis calculations based on CCSN
simulations for progenitors at the low-mass end of Fe core formation for which
the progenitor structure is similar to an ECSN progenitor (i.e., with a speep
core-density gradient; see \citealp{Mueller2016a}, their Fig.~1). The
nucleosynthesis is similar to cECSNe although there are some differences
(Fig.~\ref{fig:both_yields}), and there is slightly more variety in the
progenitor structures. Like ECSNe, these low-mass FeCCSN events are
also expected to occur in a relatively narrow mass range (less than $1\,
M_\odot$, A.  Heger, private communication).

Milky Way GCE models mixing only tECSNe and low-mass FeCCSNe with the fiducial
model (Figs.~\ref{fig:gce_both}c and d) were computed, which used the yield from
models u8.1 and z9.6 from \citet{Wanajo2018a}, respectively\footnote{The
progenitor models of e8.8 \citep{Nomoto1987, Miyaji1987}, u8.1, and z9.6
(unpublished; an extension of \citet{Heger2010a}) are those
with zero-age main-sequence masses (and initial metallicity) of $8.8\, M_\odot$
($1\, Z_\odot$), $8.1\, M_\odot$ ($10^{-4}\, Z_\odot$), and $9.6\, M_\odot$
($0\, Z_\odot$), respectively. The models u8.1 and z9.6 are at the low-mass ends
of CCSN progenitors with the adopted metallicities; the latter exhibits slightly steeper core-density gradient
\citep{Mueller2016a}. Note that all relevant isotopes ($^{48}$Ca and heavier) are
made in the innermost region of exploding material, in which the initial
metallicity has no effect (high temperature and weak interaction reset the
abundance distribution).}. The results in panels b and d look very similar
indeed, suggesting that the role of cECSNe in GCE could potentially be
superseeded by low-mass FeCCSNe. This would be the case if all ECSNe are in fact
tECSNe, which is still under debate. We concede, however, that the
model in Fig.~\ref{fig:gce_both}c performs more poorly for several of the
isotopes in the Zn--Zr region. For a good fit when there are no cECSNe, most
low-mass FeCCSNe would need to exhibit yields similar to those from the z9.6
model.

Finally, we have constructed a model in which we combine tECSNe, cECSNe, and the
two low-mass FeCCSNe (Figure~\ref{fig:gce_both}e). The rates have been tuned by
hand to bring a maximum number of isotopes close to the Solar composition,
within a factor of two. This model is not aimed to be the best-fit model, but
rather a proof of concept that ECSNe and low-mass FeCCSNe can be combined
together without creating any tension. Conversely to the models with no cECSNe,
this model requires that most low-mass FeCCSNe produce yields similar to the
u8.1 model. In this model we are able to match the solar abundance distribution
to within a factor of approximately two for almost all Zn--Zr isotopes except
for $^{84}$Sr and $^{96}$Zr. We note that neither cECSNe nor low-mass
	FeCCSNe produce a great deal of isotopes below $A = 48$ because of the
	very thin C/O shells between the cores and the low-density H/He
	envelopes of their progenitors. Similarly, the tECSNe models have such
	high production factors for $^{48}$Ca, $^{50}$Ti and $^{54}$Cr that when
	they occur at a rate that reproduces the solar inventory of $^{48}$Ca,
	very little material with $A < 48$ is produced. This is the reason why
new tensions are not introduced when including the models, they essentially just
fill in some of the missing gaps.

\begin{figure*}
	\includegraphics[width=0.98\textwidth]{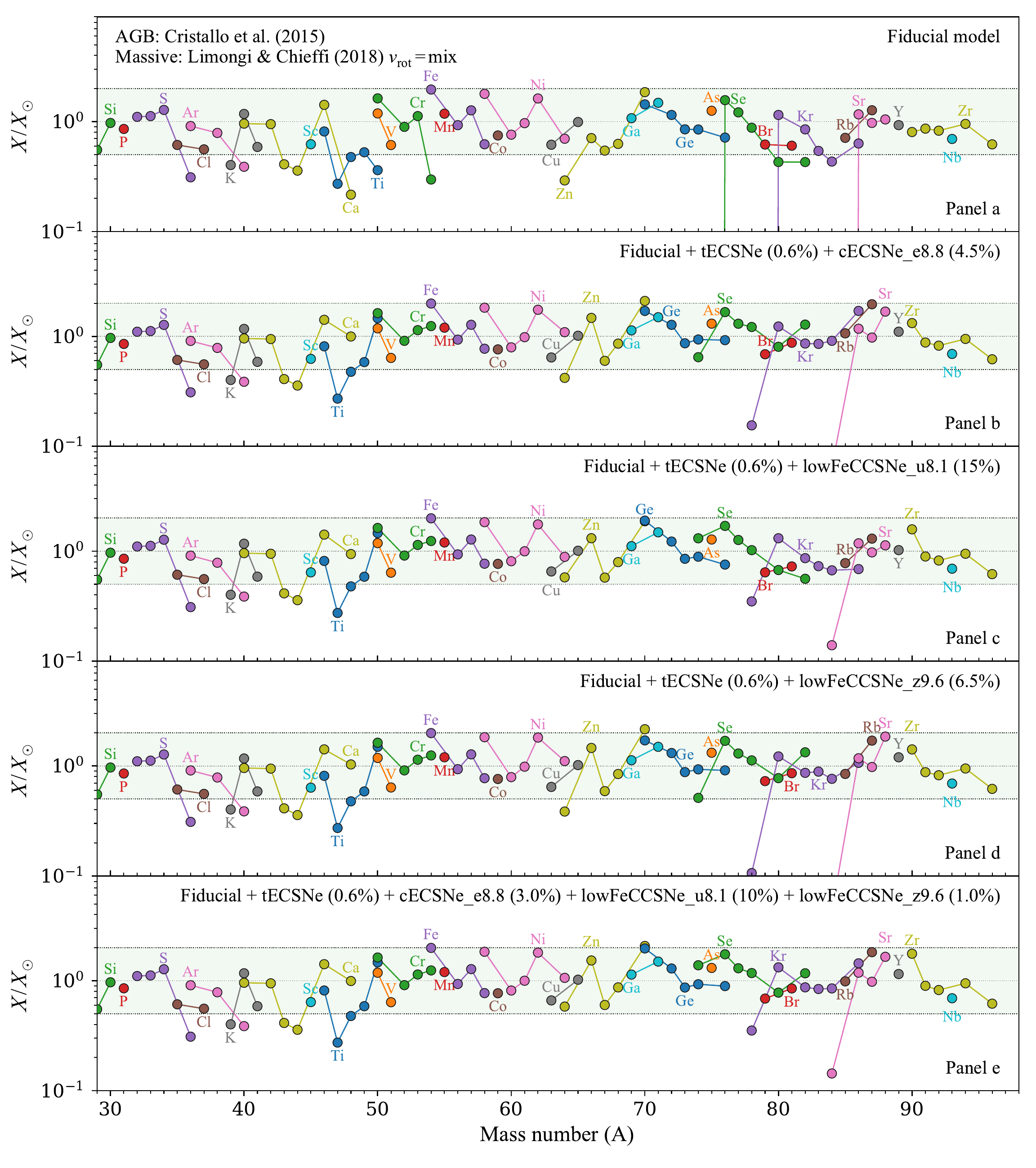}
	\caption{
	    Same as Figure~\ref{fig:gce_both}, but using the rotating massive
	    star models of  \cite{limongichieffi18}. We adopted the same
	    metallicity-dependent mix of rotation velocities as presented in
	    Figure~4 of \cite{Prantzos2018a}. The tECSNe rate has been increased
	    to 0.6\,\% to recover the abundance of $^{48}$Ca, while all other
	    rates were kept as in Figure~\ref{fig:gce_both}.
	}
	\label{fig:gce_both_rotation}
\end{figure*}

\subsubsection{Including rotating massive stars}

Figure~\ref{fig:gce_both_rotation} shows the same chemical evolution models as
in Figure~\ref{fig:gce_both}, but using the rotating models of
\cite{limongichieffi18} along with the metallicity-dependent mixture of rotation
velocities as adopted in \cite{Prantzos2018a}. Using those models instead of the
non-rotating ones required a slight re-calibration of our GCE models in terms of
gas fraction and gas outflows. To recover a similar fit for $^{48}$Ca, we
increased the tECSN rate from 0.5\,\% to 0.6\,\%. But the rates for cECSNe and
low-mass FeCCSNe are the same as in Figure~\ref{fig:gce_both}.

Because of the neutron-capture elements produced in the rotating models of
\cite{limongichieffi18}, the addition of cECSNe and low-mass FeCCSNe only
improves the predictions for a limited number of isotopes compared to when using
the non-rotating models. But our predictions still demonstrates that cECSNe and
low-mass FeCCSNe could occur individually at a rate between about 1\,\% to
10\,\% the rate of regular CCSNe without introducing any tension, besides
possibly $^{70}$Zn (see Panels b and d of Figure~\ref{fig:gce_both_rotation}).
In particular, those are needed in our models to improve the predictions for
$^{64}$Zn, $^{80,82}$Se, $^{84}$Kr, and the $p$-process isotope $^{74}$Se. In
any case, tECSNe are still needed to reproduce $^{48}$Ca, $^{50}$Ti, and
$^{54}$Cr all together.

Our predictions for the fiducial models with stellar rotation are similar to the
ones presented in \cite{Prantzos2018a} from Si to Zn. However, we predict an
underproduction of $^{50}$Ti and $^{54}$Cr compared to the latter study. This is
because we used the delayed-detonation N100 model of \cite{Seitenzahl2013a} for
Type~Ia supernovae while \cite{Prantzos2018a} used the W7 model of
\cite{iwamoto99}, which results from a one-dimensional simulation. This geometry
implies that the ashes remain at the stellar center for an artificially long
period of time being exposed to high densities.  Consequently, neutronization by
electron capture reactions is increased.  In contrast, the three-dimensional
N100 model self-consistently includes buoyancy effects, which quickly drive
burned material to low-density regions.  Moreover, the W7 yields of
\citet{iwamoto99} do not yet account for the revised electron capture rates of
\citet{langanke2000} and thus overestimate the production of $^{50}$Ti and
$^{54}$Cr \citep{brachwitz2000}. For the neutron-capture elements, the
differences likely come from the fact that we did not include the r-process in
our calculations in order to leave room for ECSNe and low-mass FeCCSNe. We note
that using rotating models solved the overproduction predicted for $^{62}$Ni.

\section{Interpretation of results}
\label{sec:interpretation}
We have demonstrated that assuming both partial thermonuclear explosion
and gravitational collapse into a NS are outcomes of ECSNe that are realised in
nature, GCE models of the Milky Way are more universally successful for
the solar inventory several isotopes including \ca, \ti,~$^{54}$Cr and
Zn--Zr. We have also shown, however, that models where all ECSNe are
thermonuclear explosions are also very successful when yields from low-mass
FeCCSNe and/or rotating massive stars are included.

Our combined model required all three types of SNe occur, and that tECSNe
(cECSNe) occur at 0.6\% (3.0\%) of the CCSN rate, and low-mass FeCCSNe occur at
11\%. This model improves the agreement with the solar abundance distribution
without introducing significant new tensions.  In this section we discuss what
the implications of such models are in a broader astrophysical context.

\subsection{Compatibility of rate estimates}

How do the assumed rates compare with other rate estimates? Single star models
\citep{Poelarends2007,Poelarends2008,Doherty2015} typically predict a range of
rates for ECSNe in the range 1--20\% of the CCSN rate. Population synthesis
simulations estimate $\sim 3\%$ \citep{Jones2019a}, with the majority coming
from binary systems but not accreting ONe WDs. Our rate for ECSNe (3.5--5.0\%)
is compatible with both these estimates when one considers the outstanding
uncertainties \citep{Jones2016a,Leung2019a}.

\subsection{Implications for ECSNe}

Since the border between gravitational collapse and thermonuclear explosion is
very sensitive to the progenitor and deflagration ignition conditions, it may
well be that some ONe core stars collapse and others explode. This requires some
variety in either the progenitors or the ignition conditions, which one might
perhaps expect if $A=24$ electron capture reactions drive convective motions in
the core \citep{Schwab2017a}.

In the model we have proposed, about 85\% of ECSNe would still produce a low
mass NS with a low kick velocity. This means that many of the phenomena
attributed to ECSNe (BeX systems with low orbital eccentricity, low kick NS
populations) could still be explained by invokation of cECSNe as their origin.

If all ECSNe were tECSNe and left behind bound WD remnants, another explanation
for low mass, low kick NSs must be sought.  It is conceivable, however, that
low-mass FeCCSNe take over the role of cECSNe completely, in which case all
ECSNe events could be thermonuclear explosions.  This implies that the lower
ECSNe rate of 0.5--0.7\% of the CCSN rate is realized. That rate could be larger
if the ejecta masses from tECSNe were smaler than current model predictions.

\section{Conclusions and outlook }

We have shown that tECSNe present a compelling explanation for the origin of
\ca, \ti~and $^{54}$Cr in the Solar System. Moreover, if tECSNe and either of
cECSNe or low-mass FeCCSNe occur in nature, GCE models of the Milky Way produce
improved results with respect to the solar abundance distribution. We note that
there is no other appreciable source known for the isotopes \ca, \ti~and
$^{54}$Cr. 

Our results add further weight to the argument that tECSNe do occur in nature.
This argument is supported by potential observations of their candidate WD
remnants \citep{Jones2019a,Raddi2019a} and isotopic ratios in pre-solar grains
\citep{Voelkening1990a,Woosley1997a,Nittler2018a,Jones2019a}. Unfortunately, all
of the evidence is circumstantial at this point.  Looking into the implications
of tECSNe for the diffuse galactic $^{60}$Fe concentration could provide further
constraints.

Including tECSNe into GCE models erases one of the last remaining blemishes in
cosmic nucleosynthesis with a viable astrophysical source model. In this work,
we have not considered a possible contribution of the r process to the Galactic
inventory of the Zn-Zr region. Further studies on the progenitor evolution
toward ECSNe, rate and explosion mechanism, and an implementation of other
possible astrophysical sources are required to substantiate this model.

\acknowledgments
The authors would like to thank the anonymous referee for their critical reading
and constructive feedback and suggestions for this manuscript.
The authors acknowledge the Lorentz Center workshop ``Electron-capture
initiated stellar collapse" at which some of the ideas in this letter
germinated.
The authors also acknowledge support from the ChETEC COST Action (CA16117)
and HITS gGmbH.
BC acknowledges support from the ERC Consolidator Grant
(Hungary) funding scheme (Project RADIOSTAR, G.A. n. 724560) and the National
Science Foundation (NSF, USA) under grant No.  PHY-1430152 (JINA Center for the
Evolution of the Elements).
This work made use of the Heidelberg Supernova
Model Archive \citep[HESMA;][]{Kromer2017a}, https://hesma.h-its.org.
The work of FR is supported by the Klaus Tschira Foundation and by the
Collaborative Research Center SFB 881 ``The Milky Way System'' of the German
Research Foundation (DFG).
This work was supported by the US Department of
Energy LDRD program through the Los Alamos National Laboratory. Los Alamos
National Laboratory is operated by Triad National Security, LLC, for the
National Nuclear Security Administration of U.S.  Department of Energy (Contract
No.  89233218NCA000001).


\software{\texttt{OMEGA+} \citep{cote18},
	Matplotlib \citep{Hunter:2007}
}


\end{document}